\documentstyle[twoside,psfig]{article}

\newcommand{\bee}{\begin{equation}}
\newcommand{\eeq}{\end{equation}}

\newcommand{\bea}{\begin{eqnarray}}
\newcommand{\eea}{\end{eqnarray}}

\catcode`\@=11
\long\def\@makefntext#1{
\protect\noindent \hbox to 3.2pt {\hskip-.9pt  
$^{{\eightrm\@thefnmark}}$\hfil}#1\hfill}		

\def\thefootnote{\fnsymbol{footnote}}
\def\@makefnmark{\hbox to 0pt{$^{\@thefnmark}$\hss}}	
	
\def\ps@myheadings{\let\@mkboth\@gobbletwo
\def\@oddhead{\hbox{}
\rightmark\hfil\eightrm\thepage}   
\def\@oddfoot{}\def\@evenhead{\eightrm\thepage\hfil
\leftmark\hbox{}}\def\@evenfoot{}
\def\sectionmark##1{}\def\subsectionmark##1{}}

\oddsidemargin=\evensidemargin
\renewcommand{\thefootnote}{\fnsymbol{footnote}}

\newcounter{sectionc}\newcounter{subsectionc}\newcounter{subsubsectionc}
\renewcommand{\section}[1] {\vspace{12pt}\addtocounter{sectionc}{1} 
\setcounter{subsectionc}{0}\setcounter{subsubsectionc}{0}\noindent 
	{\tenbf\thesectionc. #1}\par\vspace{5pt}}
\renewcommand{\subsection}[1] {\vspace{12pt}\addtocounter{subsectionc}{1} 
	\setcounter{subsubsectionc}{0}\noindent 
	{\bf\thesectionc.\thesubsectionc. {\kern1pt \bfit #1}}\par\vspace{5pt}}
\renewcommand{\subsubsection}[1] {\vspace{12pt}\addtocounter{subsubsectionc}{1}
	\noindent{\tenrm\thesectionc.\thesubsectionc.\thesubsubsectionc.
	{\kern1pt \tenit #1}}\par\vspace{5pt}}

\newcounter{appendixc}
\newcounter{subappendixc}[appendixc]
\newcounter{subsubappendixc}[subappendixc]
\renewcommand{\thesubappendixc}{\Alph{appendixc}.\arabic{subappendixc}}
\renewcommand{\thesubsubappendixc}
	{\Alph{appendixc}.\arabic{subappendixc}.\arabic{subsubappendixc}}

\renewcommand{\appendix}[1] {\vspace{12pt}
        \refstepcounter{appendixc}
        \setcounter{figure}{0}
        \setcounter{table}{0}
        \setcounter{lemma}{0}
        \setcounter{theorem}{0}
        \setcounter{corollary}{0}
        \setcounter{definition}{0}
        \setcounter{equation}{0}
        \renewcommand{\thefigure}{\Alph{appendixc}.\arabic{figure}}
        \renewcommand{\thetable}{\Alph{appendixc}.\arabic{table}}
        \renewcommand{\theappendixc}{\Alph{appendixc}}
        \renewcommand{\thelemma}{\Alph{appendixc}.\arabic{lemma}}
        \renewcommand{\thetheorem}{\Alph{appendixc}.\arabic{theorem}}
        \renewcommand{\thedefinition}{\Alph{appendixc}.\arabic{definition}}
        \renewcommand{\thecorollary}{\Alph{appendixc}.\arabic{corollary}}
        \renewcommand{\theequation}{\Alph{appendixc}.\arabic{equation}}
        \noindent{\tenbf Appendix \theappendixc #1}\par\vspace{5pt}}
\newcommand{\subappendix}[1] {\vspace{12pt}
        \refstepcounter{subappendixc}
        \noindent{\bf Appendix \thesubappendixc. {\kern1pt \bfit #1}}
	\par\vspace{5pt}}
\newcommand{\subsubappendix}[1] {\vspace{12pt}
        \refstepcounter{subsubappendixc}
        \noindent{\rm Appendix \thesubsubappendixc. {\kern1pt \tenit #1}}
	\par\vspace{5pt}}

\newcommand{\textlineskip}{\baselineskip=13pt}
\newcommand{\smalllineskip}{\baselineskip=10pt}

\def\eightcirc{
\begin{picture}(0,0)
\put(4.4,1.8){\circle{6.5}}
\end{picture}}
\def\eightcopyright{\eightcirc\kern2.7pt\hbox{\eightrm c}} 

\newcommand{\copyrightheading}[1]
	{\vspace*{-2.5cm}\smalllineskip{\flushleft
	{\footnotesize Modern Physics Letters A, #1}\\
	{\footnotesize $\eightcopyright$\, World Scientific Publishing
	 Company}\\
	 }}

\newcommand{\publisher}[2]{{\begin{center}\footnotesize\smalllineskip 
	Received #1\\
	Revised #2
	\end{center}
	}}

\def\abstracts#1#2#3{{
	\centering{\begin{minipage}{4.5in}\footnotesize\baselineskip=10pt
	\parindent=0pt #1\par 
	\parindent=15pt #2\par
	\parindent=15pt #3
	\end{minipage}}\par}} 

\renewenvironment{thebibliography}[1]
	{\frenchspacing
	 \ninerm\baselineskip=11pt
	 \begin{list}{\arabic{enumi}.}
        {\usecounter{enumi}\setlength{\parsep}{0pt}     
	 \setlength{\leftmargin 12.7pt}{\rightmargin 0pt} 
         \setlength{\itemsep}{0pt} \settowidth
	{\labelwidth}{#1.}\sloppy}}{\end{list}}

\newcounter{itemlistc}
\newcounter{romanlistc}
\newcounter{alphlistc}
\newcounter{arabiclistc}

\newcommand{\fcaption}[1]{
        \refstepcounter{figure}
        \setbox\@tempboxa = \hbox{\footnotesize Fig.~\thefigure. #1}
        \ifdim \wd\@tempboxa > 5in
           {\begin{center}
        \parbox{5in}{\footnotesize\smalllineskip Fig.~\thefigure. #1}
            \end{center}}
        \else
             {\begin{center}
             {\footnotesize Fig.~\thefigure. #1}
              \end{center}}
        \fi}

\newcommand{\tcaption}[1]{
        \refstepcounter{table}
        \setbox\@tempboxa = \hbox{\footnotesize Table~\thetable. #1}
        \ifdim \wd\@tempboxa > 5in
           {\begin{center}
        \parbox{5in}{\footnotesize\smalllineskip Table~\thetable. #1}
            \end{center}}
        \else
             {\begin{center}
             {\footnotesize Table~\thetable. #1}
              \end{center}}
        \fi}

\def\@citex[#1]#2{\if@filesw\immediate\write\@auxout
	{\string\citation{#2}}\fi
\def\@citea{}\@cite{\@for\@citeb:=#2\do
	{\@citea\def\@citea{,}\@ifundefined
	{b@\@citeb}{{\bf ?}\@warning
	{Citation `\@citeb' on page \thepage \space undefined}}
	{\csname b@\@citeb\endcsname}}}{#1}}

\newif\if@cghi
\def\cite{\@cghitrue\@ifnextchar [{\@tempswatrue
	\@citex}{\@tempswafalse\@citex[]}}
\def\citelow{\@cghifalse\@ifnextchar [{\@tempswatrue
	\@citex}{\@tempswafalse\@citex[]}}
\def\@cite#1#2{{$\null^{#1}$\if@tempswa\typeout
	{IJCGA warning: optional citation argument 
	ignored: `#2'} \fi}}

\def\pmb#1{\setbox0=\hbox{#1}
	\kern-.025em\copy0\kern-\wd0
	\kern.05em\copy0\kern-\wd0
	\kern-.025em\raise.0433em\box0}

\def\fnt#1#2{\footnotetext{\kern-.3em
	{$^{\mbox{\scriptsize #1}}$}{#2}}}

\def\fpage#1{\begingroup
\voffset=.3in
\thispagestyle{empty}\begin{table}[b]\centerline{\footnotesize #1}
	\end{table}\endgroup}

\def\runninghead#1#2{\pagestyle{myheadings}
\markboth{{\protect\footnotesize\it{\quad #1}}\hfill}
{\hfill{\protect\footnotesize\it{#2\quad}}}}
\font\tenrm=cmr10
\font\tenit=cmti10 
\font\tenbf=cmbx10
\font\bfit=cmbxti10 at 10pt
\font\ninerm=cmr9

\font\eightrm=cmr8

\def\qed{\hbox{${\vcenter{\vbox{			
   \hrule height 0.4pt\hbox{\vrule width 0.4pt height 6pt
   \kern5pt\vrule width 0.4pt}\hrule height 0.4pt}}}$}}

\renewcommand{\thefootnote}{\fnsymbol{footnote}}	

\begin{document}
\setlength{\textheight}{7.7truein}  

\runninghead{Is gravity quantum mechanical $\ldots$}{ Dynamics of Gravity from  $\ldots$}

\normalsize\textlineskip
\thispagestyle{empty}
\setcounter{page}{1}

\copyrightheading{}			

\vspace*{0.88truein}

\fpage{1}
\centerline{\bf  Is gravity an intrinsically  quantum phenomenon ?}
\centerline{\bf  Dynamics of Gravity from the Entropy of Spacetime}
\centerline{\bf and  the Principle of Equivalence}
\baselineskip=13pt
\vspace*{0.37truein}
\centerline{\footnotesize T. PADMANABHAN}
\baselineskip=12pt
\centerline{\footnotesize\it IUCAA, P.O.Box 4, Ganeshkhind,}
\baselineskip=10pt
\centerline{\footnotesize\it Pune 411 007, Maharashtra, India.}

\vspace*{0.225truein}

\publisher{(received date)}{(revised date)}

\setcounter{footnote}{0}
\renewcommand{\thefootnote}{\alph{footnote}}

\vspace*{0.21truein}
\abstracts{The two surprising features of gravity are (a) the principle of equivalence and
(b) the connection between gravity and thermodynamics. 
 Using principle of equivalence
    and special relativity in the {\it local inertial frame}, one could obtain the insight that gravity
    must possess a geometrical description. I show that, using the same principle of equivalence,
    special relativity  and quantum theory in the {\it local 
    Rindler frame} one can obtain the Einstein-Hilbert action
functional for gravity and thus the dynamics of the spacetime.  This approach, which essentially 
  involves postulating that
the horizon area must be proportional to the entropy, uses the 
 local Rindler frame  as a natural extension of the 
local inertial frame
and leads to the interpretation that the gravitational action represents the free energy of the 
spacetime geometry.
 As an aside, one also obtains  a natural explanation
as to:
(i) why the covariant action for gravity contains second derivatives of the metric tensor and 
(ii)  why the   gravitational coupling constant is positive. 
The analysis suggests that gravity is intrinsically
holographic and even intrinsically quantum mechanical. }{}{}


\vspace*{1pt}\textlineskip

\def\rb{\right)}
\def\lb{\left(}

\def\bld#1{{\bf #1 }}

    \section{A possible synthesis: Motivation and summary}
    
    Two aspects of gravity stand out among the rest as the most surprising: (i) the
existence of the principle of equivalence and (ii) the connection between gravity and thermodynamics.
    The   principle of equivalence ---  in the broadest sense ---
    postulates that physical phenomena taking place around any event
    ${\cal P}$  in  a local region of spacetime   cannot be distinguished from the corresponding phenomena 
     taking place in a suitably chosen non inertial frame. This 
     principle finds its natural expression when gravity is described as a manifestation of
     curved spacetime. The second surprise regarding gravity is the deep connection
     it has with thermodynamics (for a review, see references [1], [7]). 
     Gravity is the only interaction which is capable
     of wrapping up regions of spacetime so that information from one region is not 
     accessible to observers at another region. Given the fact that entropy of systems
     is closely related to accessibility of information, it is inevitable that there will be
    some connection between gravity and thermodynamics. But, in contrast to the
principle of equivalence,  years of 
    research in this field (see, for a sample of references and related work [2]),
    has not 
  led to  something more profound or fundamental arising out of this feature. 
    
    It is possible to learn a lesson from the way Einstein handled the principle of 
    equivalence and  apply it in the context of the connection between thermodynamics
    and gravity. Einstein did not attempt to ``derive" principle of equivalence in the conventional 
    sense of the word. Rather, he accepted it as a key feature which must find expression in the 
    way gravity is described --- thereby obtaining  a geometrical description of gravity. 
    Once the geometrical interpretation of gravity is accepted, it follows that there will arise
    surfaces which act as one way membranes for information and thus will lead to some connection
    with thermodynamics. It is, therefore, more in tune with the spirit of Einstein's analysis
    to {\it accept} an inevitable connection between gravity and thermodynamics and ask 
    what such a connection would imply. 
    Let me elaborate this idea further in order to show how powerful it is. 
    
    The first step in the logic, as indicated above, is the principle of equivalence, which finds
    its  natural expression in a model for gravity with the metric tensor of the spacetime
    $g_{ab}$ being the fundamental variable. 
    This allows one to define a coordinate system around any event ${\cal P}$ in a region of size
    $L$ (with $L^2(\partial^2 g/g)   \ll 1$ but $L(\partial g/g) $ being arbitrary, where
    $\partial^n g$ denotes the typical value of the $n$th derivative of the metric tensor
     at ${\cal P}$)
    in  which the spacetime is locally inertial.
    Using the laws of special relativity in this locally inertial frame and expressing them
    in a generally covariant manner (using  the ``comma-to-semicolon rule'', say) one can 
    describe the coupling of gravity to other matter fields.

    As the second step, we want to give expression to the fact that there is a deep connection between
    one way membranes arising in a spacetime and thermodynamical entropy. This, of course,
    is not possible in the local inertial frame since the quantum field theory in that frame, say, 
    does not recognize any non trivial geometry of spacetime. But it is possible to achieve our aim
    by using a uniformly accelerated frame around ${\cal P}$. In fact, 
    around  any event ${\cal P}$ we have fiducial observers anchored
    firmly   in space with ${\bf x} = $ constant and  the four-velocity $u^i = g_{00}^{-1/2}(1,0,0,0)$
    and  acceleration $a^i = u^j \nabla_j u^i$. This allows us to define a second natural
    coordinate system around any event by using the Fermi-Walker transported coordinates
    corresponding to these accelerated observers. I shall call this the local Rindler frame.
    [Operationally, this coordinate system is most easily constructed by first transforming to
    the locally inertial frame and then using the standard transformations between the 
    inertial coordinates and the Rindler coordinates.]
    This local Rindler frame will lead to a natural notion of horizon and associated temperature.
    The key new idea in my approach will be to postulate that the horizon in the 
    local Rindler frame also has an entropy per unit transverse area and  demand that any
    description of gravity must have this feature incorporated in it. 
    
    What will such a postulate lead to? {\it Incredibly enough, it leads to the correct Einstein-Hilbert
    action principle   for gravity}. Note that the original approach of Einstein making use of the principle of
    equivalence lead only up to the {\it kinematics} of gravity --- viz., that gravity is described by a curved
    spacetime with a non trivial metric $g_{ab}$ --- and cannot tell us how the {\it dynamics} of the 
    spacetime is determined. Taking the next step, using the local Rindler frame and demanding that
    gravity must incorporate the thermodynamical aspects lead to the action functional itself.

    This itself is interesting; but I will show that this approach also throws light
    on what has been usually considered a completely different issue: Why does
    the Einstein-Hilbert action contain second derivatives of the metric tensor?
    One must admit that this is peculiar --- in the sense that no other interaction
    has this feature. It is also closely related to the geometrical nature of gravity
    which  prevents the construction of a covariant scalar which is quadratic in the
    first derivatives of gravity. The analysis presented in this paper ``builds up''
    the Einstein-Hilbert action from its surface behaviour and, in this sense,
    shows that gravity is intrinsically holographic.\cite{grf} 
    In the literature, the term `holographic' is used with different meanings \cite{holography};
     I use this term with the specific meaning
    that given the form of the action on a two dimensional surface, there is a way of obtaining
    the full bulk action. In fact, this can be done for any theory but it finds a natural place
    only in the case of gravity. 

In the $(3+1)$ formalism, this leads to the interpretation of the gravitational action as the
free energy of spacetime. Einstein's equations are equivalent to the principle of minimization of 
free energy in thermodynamics.
    
    This approach opens up a new point of view regarding gravity. Using principle of equivalence
    and special relativity in the local inertial frame, one could obtain the insight that gravity
    must possess a geometrical description. Using the same principle of equivalence,
    special relativity {\it and quantum theory in the local 
    Rindler frame one can obtain the dynamics of the spacetime}. The key input is the connection
    between entropy and the horizon area and this requires introduction of a length scale into the
    problem which needs to be determined from observations. With hindsight, we know that this 
    length scale will be proportional to the Planck length $L_P =(G\hbar/c^3)^{1/2}$. But in the
approach advocated here,
   it is more natural to write the Newtonian gravitational force as $F = (c^3L_P^2/\hbar)(m_1m_2/r^2)$
    suggesting that gravity is intrinsically quantum mechanical. 
    The broader implications are discussed in the last section.
    
    \section{Action functionals with second derivatives}
    
    As described in the previous section, I will introduce a  new postulate  which relates
    the transverse spatial area of the horizon in the local Rindler frame to the entropy
   of the horizon. Since the entropy can be related to the Euclidean action
    we will be able to determine the form of the action. It turns out, however, that it is slightly
easier to reach this goal if I start with some 
    general results regarding  action functionals and then connect it up with the problem
    at hand. This is important because some of these results do not seem to 
    received adequate attention in the  literature.
    
    Consider a physical system described by the  dynamical variables $q$ 
    and a Lagrangian $L(\partial q,q)$ made of the dynamical variables and their
    first derivatives. The ideas described below work for any number of variables
    (so that $q $ can be a multicomponent entity) dependent on space and time.
    But I shall illustrate the idea first in the context of point mechanics.
    Given any Lagrangian $L(\partial q, q)$ involving only up to the first derivatives of the
     dynamical variables, it is 
         {\it always} possible
to construct another Lagrangian $L'(\partial^2q,\partial q,q)$, involving 
second derivatives such that it describes the same dynamics \cite{tpdlb}. The prescription is:
\begin{equation}
L'=L-{d\over dt}\left(q{\partial L\over \partial\dot q}\right)
\label{lbtp}
\end{equation}
While varying the $L'$, one keeps the {\it momenta} $(\partial L/\partial\dot q)$ fixed
at the endpoints rather than $q'$s. 
This is most easily seen by explicit variation; we have
   \begin{eqnarray}
 \delta A' &= &    \int\limits^{{\cal P}_2}_{{\cal P}_1} dt \left[ {\partial L \over \partial  q }\delta q +  {\partial L \over \partial \dot q }\delta \dot q \right] - \delta \lb q {\partial L \over \partial \dot q } \rb \bigg \vert^{{\cal P}_2}_{{\cal P}_1}\nonumber \\
&=& \int\limits^{{\cal P}_2}_{{\cal P}_1}dt \left[ {\partial L \over \partial  q } - {d \over dt} \lb {\partial L \over \partial \dot q } \rb \right] \delta q + \delta q \lb {\partial L \over \partial \dot q } \rb \bigg \vert^{{\cal P}_2}_{{\cal P}_1} - \delta q 
\lb {\partial L \over \partial \dot q } \rb\bigg\vert^{{\cal P}_2}_{{\cal P}_1} - q \delta  \lb {\partial L \over \partial \dot q } \rb \bigg \vert^{{\cal P}_2}_{{\cal P}_1} \nonumber \\
&= &\int\limits^{{\cal P}_2}_{{\cal P}_1}  dt \left[ {\partial L \over \partial  q } - {d \over dt} \lb {\partial L \over \partial \dot q } \rb \right] \delta q - q \delta p \bigg \vert^{{\cal P}_2}_{{\cal P}_1}
\end{eqnarray}
If we keep $\delta p = 0$ at the end points while varying $L'$, then we get back the same
Euler-Lagrange equations  as obtained by varying $L$ and keeping $\delta q =0$ at end points.
Since $L = L \lb \dot q, q \rb$, the quantity $q \lb \partial L /\partial \dot q \rb$ will also depend on $\dot q $. 
So the term $d \lb q \partial L / \partial \dot q \rb / dt $ will involve $\ddot q$. Thus $L'$ contains second derivatives of $q$ while $L$ contains only up to first derivatives. In spite of the fact that $L'$ contains second derivatives of $q$, the equations of motion arising from $L'$ are only second order for variation
with $\delta p =0$ at end points.
It can be shown that, in the  path integral formulation of quantum theory,  the modified Lagrangian $L'$ 
correctly describes the transition amplitude between states with given momenta
(see p. 170 of  [5]).

In the case of flat spacetime quantum field theory of a scalar field, say, we can start with a Lagrangian
density of the form $L(\partial\phi,\phi)$ and obtain $L'(\partial^2\phi,\partial\phi,\phi)$ by a similar procedure.
In terms of the actions, the relation will be
\begin{equation}
A' = \int_{\cal V} d^4 x\, L' (\partial^2 \phi, \partial\phi,\phi) =  \int_{\cal V} d^4 x \, L( \partial\phi,\phi) -
    \int_{\cal V} d^4 x\  \partial_a \left[ \phi {\partial L\over \partial(\partial_a \phi)}\right]\equiv A - {\cal S}
    \label{phiA}
    \end{equation}
    The second term ${\cal S}$ can, of course, be converted into a surface integral over the 3-dimensional
    boundary $\partial {\cal V}$. If we consider a static field configuration (in some Lorentz frame)
    then the second term in (\ref{phiA}) will have the integrand $\nabla\cdot[\phi(\partial L/\partial(\nabla \phi)]$
    which can be converted to an integral over a {\it two} dimensional surface on the boundary
    $ \partial \partial {\cal V}$.  Taking the time integration over an interval $(0,T)$, the second term
    in (\ref{phiA}), for static field configurations, will reduce to 
    \begin{equation}
    {\cal S} =  \int_0^T dt \int_{\partial{\cal V}} d^3 x \, \nabla \cdot \left[ \phi {\partial L\over \partial(\nabla \phi)}\right]
    =T \int_{\partial\partial{\cal V}} d^2 x\  \hat{\bf n}\cdot \phi{\partial L\over \partial(\nabla \phi)}
    \equiv  \int_{\partial\partial{\cal V}} d^2 x \hat{\bf n}\cdot {\bf P}
    \label{surface}
    \end{equation}
    This procedure allows one to reconstruct the bulk action if the surface term is known.
    As an example, let us assume the surface term has the above form with ${\bf P} =- (T/2) \nabla(\phi^2)$.
    This shows that $[\partial L/\partial(\nabla \phi)] =-\nabla \phi$ leading to 
    $L = -(1/2) (\nabla \phi)^2-V(\phi) $ (where $V$ is an arbitrary scalar function
of $\phi$) for static field configurations. Lorentz invariance now allows us to
    fix the time dependence leading to $ L =(1/2) [\partial_a \phi\partial^a\phi]-V$.
    This is the standard  first order Lagrangian; adding the surface term will give the second order Lagrangian
    for a scalar field to be $L' = - \phi \Box \phi - (1/2) [\partial_a \phi\partial^a\phi]-V$. 
    
    While
    this procedure is viable and consistent, it is not of much use in this context and, more importantly,
    suffers from the following serious flaws: (i) There is no real reason or motivation to believe that the 
    ${\bf P}$ in the surface term in (\ref{surface}) 
    should be of the form $- (T/2) \nabla(\phi^2)$. 
    (ii) The scaling with respect to $T$ in the postulated surface term is strange 
    and there is no natural value for $T$. 
    (iii) The Lagrangian $L'$ which one obtains by this method is in no way superior to the 
    standard Lagrangian $L$. Because of these reasons, 
     one might as well postulate the original Lagrangian
    rather than obtain it in such a convoluted way.

    I will now show that the situation is
     very different in the case of gravity and a similar procedure appears to be
    natural and logical at every step of the way. 
    
    \section{Gravitational dynamics from spacetime thermodynamics}
    
    After the warm up regarding action functionals containing second derivative of dynamical
    variables, let us now proceed to implement the basic idea introduced in section 1.
    The principle of equivalence leads to a geometrical description of gravity in which
    $g_{ab}$ are the fundamental variables. We expect the dynamics of gravity to be described by some
    {\it unknown} action functional 
    \begin{equation}
    A =
     \int d^4x \sqrt{-g} L(g,\partial g) \equiv \int d^4 x \sqrt{-g} L(g, \Gamma)
    \label{aquad}
    \end{equation}
    involving $g_{ab}$s and their first derivatives $\partial_c g_{ab}$ or, 
    equivalently, the set $[g_{ab}, \Gamma^i_{jk}]$ where $\Gamma$s are the standard 
    Christoffel symbols. 
    From the discussion in the last section, it is obvious that the {\it same}  equations
    of motion can be obtained from another (as yet unknown) action:
    \begin{eqnarray}
    A' &=& \int d^4x \sqrt{-g} L - \int d^4x \partial_c \left[ g_{ab}
     {\partial \sqrt{-g} L \over \partial(\partial_c g_{ab})}
    \right]  
   \equiv \int d^4x \sqrt{-g} L - \int d^4x \partial_c \left[ g_{ab} \pi^{abc}
    \right]\nonumber\\
    & \equiv& A - \int d^4 x \partial_c (\sqrt{-g}V^c)\equiv A - \int d^4 x \partial_c P^c 
    \label{aeh}
    \end{eqnarray}   
    where $V^c $ is made of $g_{ab} $ and $\Gamma^i_{jk}$. Further, $V^c$ must be linear
    in the $\Gamma$'s since the original Lagrangian $L$ was quadratic in the first derivatives
    of the metric. 
    Since $\Gamma$s vanish in the local inertial frame and the metric reduces to
    the Lorentzian form, the action $A$ cannot be generally covariant.
    However, the action $A'$ involves the second derivatives  of the metric and hence
    we shall demand that the action $A'$ must be generally covariant.

    To obtain a quantity, which is linear in $\Gamma$s
     and having  a single index $c$, from $g_{ab} $ and $\Gamma^i_{jk}$,
     we must contract on two of the indices on $\Gamma$
    using the metric tensor. Hence the most general choice for $V^c$ is the linear combination
      \begin{equation}
    V^c =  \left(a_1 g^{ck} \Gamma^m_{km} +a_2 g^{ik} \Gamma^c_{ik}\right) 
    \label{defvc}
     \end{equation}
     where $a_1$ and $a_2$ are numerical constants. Using the identities $\Gamma^m_{km} =\partial_k
     (\ln \sqrt{-g})$, \ $\sqrt{-g}g^{ik}\Gamma^c_{ik} = -\partial_b(\sqrt{-g}g^{bc})$,
     we can rewrite the expression for $P^c \equiv \sqrt{-g}V^c$ as 
     \begin{equation}
    P^c =\sqrt{-g}V^c=
    c_1g^{cb} \partial_b \sqrt{-g} +c_2 \sqrt{-g} \partial_b g^{bc}
    \label{defpc}
    \end{equation}
    where $c_1\equiv a_1 - a_2,\  c_2\equiv -a_2$ are two other numerical constants. 
    If we can fix these coefficients by using a physically well motivated prescription, then
    we can determine the surface term and by integrating, the Lagrangian $L$.
    I will now show how this can be done.
    
    Let us consider a static spacetime in which all $g_{ab}$s are
    independent of $x^0$ and $g_{0\alpha} =0$. Around any given event ${\cal P}$
    one can construct a local Rindler frame with an acceleration 
    of the observers with ${\bf x} $ = constant, given by $a^i = (0,{\bf a})$ and
    ${\bf a}= \nabla (\ln \sqrt{g_{00}})$.
    This Rindler frame will have a horizon which is a plane surface normal to the 
    direction of acceleration and a temperature $T=|{\bf a}|/2\pi$ associated with this horizon.
     I shall postulate that the entropy associated with
    this horizon is proportional to its area or, more precisely,,  
    \begin{equation}
    {dS\over dA_\perp} = {1\over {\cal A}_P}
    \label{postulate}
    \end{equation}
    where ${\cal A}_P$ is a fundamental constant with the dimensions of area. 
    It represents the minimum areas required to hold unit amount of  information 
     and our postulate demands that this number be finite.
    Given the temperature of the horizon, one can construct a canonical ensemble 
    with this temperature and relate the Euclidean action to the thermodynamic entropy
(see, e.g, [6],[7]).
    Since the Euclidean action can be interpreted as the entropy in the canonical
    ensemble, I will demand that the quadratic action $A$ in equation (\ref{aquad})
    should be related to the entropy by  $S = -A$ (with the minus sign arising from standard 
    Euclidean continuation \cite{tplongpap}), when evaluated in the local 
    Rindler frame with the temperature $T$.
    I next note that the action $A'$ is generally covariant and hence will vanish
    in the local Rindler frame. It follows that the numerical value of the action  $A$ in the local
    Rindler frame is the same as the surface term in equation (\ref{aeh}).
     That is, in the static local Rindler frame, we have 
     \begin{equation}
     A  = \int_{\partial {\cal V}} d^4 x \partial_c P^c = \int_0^\beta dt \int_{\partial {\cal V}} d^3 x
     \nabla \cdot {\bf P} = \beta \int_{\partial\partial {\cal V}} d^2 x_\perp \hat{\bf n}\cdot {\bf P}
     \label{keyeqn}
     \end{equation}
     I  have restricted the time integration to an interval $(0,\beta)$
     where $\beta = (2\pi /|{\bf a}|) $ is the inverse temperature in the Rindler frame.
     This is {\it needed} since the Euclidean action will be periodic in the imaginary
     time with the period $\beta$. 
    We shall choose the local Rindler frame such that the acceleration is along
    the $x^1=x$ axis, say, and the line element to be 
    \begin{equation}
    ds^2 = (1+ ax)^2 dt^2 - d{\bf x}^2
    \end{equation}
    Evaluating $P^a$ using equation (\ref{defpc})  we find that only the $c_1$ term
    contributes and $P^a=(0,-c_1a, 0, 0)$ so that the action in (\ref{keyeqn})
    becomes (on using $a\beta=2\pi$)
    \begin{equation}
    A= -c_1 a \beta \int d^2 x_\perp = - 2\pi c_1  \int d^2 x_\perp= -2\pi c_1 A_\perp
    \label{detcone}
    \end{equation}
    where $A_\perp$ is the transverse area of the $(x^2-x^3)$ plane.
    The entropy $S$, which is equal the Euclidean action, will be given by
    $S = -A$ with the minus sign arising from standard Euclidean continuation.
    Invoking our postulate (\ref{postulate}) we find that $c_1 = (2\pi {\cal A}_P)^{-1}$.
    
    To fix $c_2$, we shall repeat the same analysis in a different coordinate system
    representing the local Rindler frame in which the line element is given by
    \begin{equation}
    ds^2 =(1+2ax) dt^2 - {dx^2\over (1+2ax)} - dx_\perp^2
    \end{equation}
    In this case, only the $c_2 $ term in (\ref{defpc}) contributes
    giving again $P^a=(0,-2ac_2, 0, 0)$.
    The action now becomes 
       \begin{equation}
    A= -2ac_2  \beta \int d^2 x_\perp = - 4\pi c_2  \int d^2 x_\perp= -4\pi c_2 A_\perp
    \label{detctwo}
    \end{equation}
    By the same argument as before, we get $c_2 = (4\pi {\cal A}_P)^{-1}$.
    Hence $P^c$ has the form 
    \begin{equation}
    P^c =   {1\over 4\pi {\cal A}_P} \left( 2g^{cb} \partial_b \sqrt{-g} + \sqrt{-g} \partial_b g^{bc}\right)
    ={\sqrt{-g}\over 4\pi {\cal A}_P}  \left( g^{ck} \Gamma^m_{km} - g^{ik} \Gamma^c_{ik}\right) 
    \label{pcfix}
    \end{equation}
    [The second equality is obtained by using the standard identities mentioned after 
    equation (\ref{defvc}).]
    This result is remarkable and let me discuss it before proceeding further.
    
    The general form of $P^c$ which we obtained in (\ref{defpc}) is not of any use unless
   we can fix  $(c_1,c_2)$. For static configurations, we can convert the extra term to an integral over time 
   and a two-dimensional spatial surface. This is in general true, as we saw in the case of the scalar field
   earlier in section 2. But in general, the result will not have any simple form and will involve an undetermined
   range of integration over time coordinate (like the factor $T$ we found in the case of the scalar field). 
   But in the case of gravity, three natural features conspire together to give an elegant form to this
   surface term. First is the fact that Rindler frame has a periodicity in Euclidean time and the range
   of integration over the time coordinate is naturally restricted to the interval $(0,\beta) = (0,2\pi/a)$.
   The second is the fact that the integrand $P^c$ is linear in the acceleration $a$ thereby neatly
   canceling with the $(1/a)$ factor arising from time integration. Finally, there are two natural
   coordinate systems for the Rindler spacetime which allows us to determine the constants
   $c_1$ and $c_2$.  It will be incredible if all these are accidental and are  not of any fundamental 
   significance. 
   (To be absolutely rigorous, we should verify that the form of $P^c$ determined in (\ref{pcfix})
   will correctly reproduce the entropy as $(A_\perp/{\cal A}_P)$  in any other coordinate
   system. One can show by a fairly straightforward analysis that this is indeed true in any
   other static coordinates in the Rindler frame with the periodicity in Euclidean time.)
   
    Given the form of $P^c$ we need to solve the 
   equation   
    \begin{equation}
 \left({\partial \sqrt{-g}L\over\partial g_{ab,c}}g_{ab}\right)=
 P^c= {1\over 4\pi {\cal A}_P} \left(2g^{cb}\partial_b\sqrt{-g} + \sqrt{-g}\partial_bg^{cb}\right)
 \label{dseq}
\end{equation}
to obtain 
the first order Lagrangian density.
    This is straightforward and we get 
\begin{equation}
\sqrt{-g}L  \equiv {1\over 4\pi {\cal A}_P} \left( \sqrt{-g}{\cal G}\right) = 
 {1\over 4\pi {\cal A}_P} \left(\sqrt{-g} \, g^{ik} \left(\Gamma^m_{i\ell}\Gamma^\ell_{km} -
\Gamma^\ell_{ik} \Gamma^m_{\ell m}\right)\right).
\label{ds}
\end{equation}  
  (It is easier to verify the result by calculating the derivative of (\ref{ds}) with respect to
  $g_{ab,c}$ and comparing with (\ref{dseq}) written in terms of Christoffel symbols.)
    This is the second surprise. The Lagrangian which we have obtained is precisely
    the first order Dirac-Schrodinger Lagrangian for gravity (usually called the $\Gamma-\Gamma$
    Lagrangian).
    Note that we have obtained it without introducing the curvature tensor anywhere in the picture.
    Once again, this is unlikely to be a mere accident. 
    
    Given the two pieces, the final second order Lagrangian follows from our equation (\ref{aeh})
    and, of course, it is the standard Einstein-Hilbert Lagrangian.    
   \begin{equation}
  \sqrt{-g} L_{grav}=\sqrt{-g}L - {\partial\over\partial x^c} \left(g_{ab}
      \pi^{abc}\right) =  \left({1\over 4\pi {\cal A}_P}\right)R\sqrt{-g}.
      \label{lgrav}
       \end{equation}
    Thus our full second order Lagrangian  {\it turns out} to be the standard 
Einstein-Hilbert Lagrangian.
The surface terms dictate the form of the Einstein Lagrangian in the bulk.
That is, the postulate of entropy  being  proportional to  the area of the horizon,
added to the requirement of general covariance, uniquely determines the gravitational action principle.  
The idea that surface areas  encode  bits of information per quantum of area  allows one to determine the nature of gravitational interaction on the bulk, which is an interesting realization of the holographic principle. 
I will conclude this section with a set of technical comments and discuss the implications of this result in the 
next section. 

(i) Since $\pi^{abc}$ only depends on the derivative of the Lagrangian
  with respect to $g_{ab,c}$, it does not change if a 
scalar function of $g_{ab}$ is added to $L$ or 
  to ${\cal G}$ in equation (\ref{ds}). Any such invariant scalar, built purely from $g_{ab}$, 
must be a constant. This implies that we can add to $R$
  an undetermined constant in (\ref{lgrav}); thus a cosmological constant
  is still allowed and --- unfortunately --- this approach cannot
  say anything about cosmological constant. Also note that our approach provides, even classically, a new route
to Einstein's theory. In (\ref{defvc}), one can pull out one overall constant and keep the ratio 
$\lambda\equiv(a_1/a_2)$ as unknown. Integrating (\ref{dseq}) and adding to the surface term will now lead to a messy
expression which can be expressed as a linear combination of $R$ and a non-covariant term multiplied by
$(1+\lambda)$. The vanishing of the non-covariant term will require  $\lambda\equiv(a_1/a_2)=-1$ and $a_2$ will remain undetermined.
This, while pedagogically interesting, does not lead to anything new.

(ii) The approach leads to new insights in the $(3+1)$ formalism of gravity as well. 
 If we foliate the spacetime by a series
of spacelike hyper-surfaces  ${\cal S}$ with $u^i$ as normal, then $g^{ik}=h^{ik}+u^iu^k$ where $h^{ik}$ is the induced metric on ${\cal S}$. 
Given the covariant derivative $\nabla_i u_j$ of the normals to ${\cal S}$, one can construct only  three vectors
$(u^j\nabla_j u^i, u^j\nabla^i u_j, u^i\nabla^j u_j)$ which are linear in covariant derivative operator. The first one is
the acceleration $a^i=u^j\nabla_j u^i$; the second identically vanishes since $u^j$ has unit norm; the third,
$u^i K$, is
proportional to the trace of the extrinsic curvature $K=\nabla^j u_j$ of ${\cal S}$. Thus 
$V^i$ in the surface term  must be  a linear combination of $u^i K$ and $a^i$ and the corresponding
term in the action must have the form 
\begin{equation}
A_{\rm surface} = \int d^4 x \, \sqrt{-g} \nabla_i \left[ \lambda_1 K u^i + \lambda_2 a^i\right]
\label{threedaction}
\end{equation}
where $\lambda_1$ and $\lambda_2$ are numerical constants. (This result is known in the conventional
$(3+1)$ formalism; see e.g., equation (21.88) of the first reference in [8] ).
Since $u^ia_i=0$, the spacelike boundaries at $x^0=$ constant gets contribution only
from $K$ while the time like surfaces like $x^1=$ constant gets contribution from the normal component
of the acceleration $a^i\hat n_i$ where $\hat n_i$ is the normal to the time like surface. 
For static spacetimes with a horizon, the time integration can be limited to the 
range ($0,\beta$) and 
 $\nabla_i a^i$ becomes $ \nabla_\alpha a^\alpha$. So the second term can be converted into
an integral  of the normal component of the acceleration $a^\alpha\hat n_\alpha$ over a {\it two} surface.
Using $(a^\alpha\hat n_\alpha) \beta = 2\pi$, we find that this term is proportional to the transverse
area and gives the entropy of the spacetime. Proceeding as before, one  would like to 
obtain a first order Lagrangian whose derivative with respect to the dynamical variables
will contribute to the surface term in (\ref{threedaction}). In the $(3+1)$ formalism,
the spatial components of the metric are the natural dynamical variables.
Requiring that the first term $\lambda_1 K u^i$  in (\ref{threedaction})
arises through our prescription will lead to the following first order action
\begin{equation}
A_I = \int d^4x\, \sqrt{-g} L =\int d^4x\, \sqrt{-g} \left[ \alpha_1 K^{\mu \nu}K_{\mu\nu} +  \alpha_2 K^2  + F({}^3g) \right] 
\label{aone}
\end{equation}
where $ \alpha_{1,2} $ are numerical constants with $(\alpha_1 +3\alpha_2) = \lambda_1$
and $F({}^3g)$ is a function of the 3-geometry. Adding (\ref{aone}) and (\ref{threedaction}) will give the full
bulk action with three undetermined numerical constants and one unknown function $F$. 
Demanding general covariance for this expression will allow us, after fairly detailed algebra,
to determine $F$ to be the scalar curvature ${}^3R$ of the 3-geometry and the relative
values of the numerical constants. This will lead to the standard ADM form of the action functional.
In the Euclidean sector, the ADM Lagrangian becomes the Hamiltonian because of the standard 
sign change in terms quadratic in  $K^{\alpha\beta}$.  The full Euclidean  action can be written in a suggestive
form as 
\begin{equation}
A_{\rm grav} = A_{\rm surface} + A_I = S - \beta H 
\end{equation}
for any spacetime geometry having a periodicity $\beta$ in the Euclidean time.
In this spirit, gravitational action represents the free energy of the spacetime; the first order term
gives the Hamiltonian (in the Euclidean sector) and the surface term gives  the entropy.

(iii) It is actually possible to motivate the form of (\ref{defvc}) directly from the analysis in the local Rindler frame.
To see this note that (\ref{keyeqn}) along with (\ref{postulate}) requires the spatial components $V^\alpha$
to be proportional to the acceleration $a^\alpha=\Gamma^\alpha_{ik}u^iu^k$.
Writing,   $a^\alpha=\Gamma^\alpha_{ik}(g^{ik}-h^{ik})$
the first term leads directly to the second term in (\ref{defvc}) while an analysis of the second
term in the local Rindler frame will give raise to the first term of (\ref{defvc}).

 \section{Conclusions}
  
  Let us take stock of the new features in this particular approach and see where it leads
  further. To begin with, we recall that this approach is a natural extension of the original
  philosophy of Einstein; viz., to use non inertial frames judiciously to understand the 
  behaviour of gravity. In the original approach, Einstein used the principle
  of equivalence which leads naturally to the description of gravity  in terms of the 
  metric tensor. Unfortunately, {\it classical} principle of equivalence cannot take us any further
  since it does not encode information about the curvature of spacetime. However, the true
  world is quantum mechanical and one would like to pursue the analogy between non inertial
  frames and gravitational field into the quantum domain. Here the local Rindler frame arises
  as the natural extension of the local inertial frame and the study of the thermodynamics of the 
  horizon shows a way of combining special relativity, quantum theory and physics in the 
  non inertial frame. I have shown that these components are adequate to determine the
  action functional for gravity and, in fact, leads very naturally to the Einstein-Hilbert action.
  
  All along the analysis, one cannot but notice the natural manner in which different 
  factors blend to give meaningful results. I started with a  prescription
  for describing any theory with first order Lagrangian in terms of another Lagrangian
  with second derivative terms. This prescription is quite general but it is pretty much 
  useless in all cases except in gravity ! We saw that the Lagrangian for a scalar field,
  say, obtained by this prescription is in no way preferable to the standard quadratic 
  Lagrangian for the scalar field. But in the case of gravity, the second order Lagrangian
  {\it turns out} to be the Einstein-Hilbert Lagrangian. This is remarkable because
  we did not introduce the curvature of spacetime explicitly into the discussion and 
  --- in fact --- the analysis was done in a local Rindler frame which is just
  flat spacetime. The idea works because the action for gravity splits up into
  two natural parts {\it neither} of which is generally covariant. Though the sum of the
  two parts (which is generally covariant) is zero, the expression for individual parts can be ascertained in the 
  local Rindler frame specifically because these parts are {\it not} generally covariant.
  
  The analysis determines the action for gravity to within an undetermined multiplicative
  factor which we have called ${\cal A}_P$. In equations (\ref{detcone}), (\ref{detctwo})
  the product $a\beta$ has the value $2\pi (c/\hbar)$ in normal units
  if the temperature is measured in energy units. Hence the two constants we determined
  will scale as $c_1 = (\hbar/2\pi c {\cal A}_P), \  c_2 = (\hbar/4\pi c {\cal A}_P)$.
  The gravitational interactions are therefore determined by the coupling
  constant $c{\cal A}_P / \hbar$. While the existence of $c$ in the gravitational
  Lagrangian is completely understandable and arises from the metric
  having, say, a $c^2 dt^2$ term, the existence of $\hbar $ in gravity
  requires further thought. One could, of course, trivially redefine ${\cal A}_P$
  as something like $\kappa \hbar$ cancel out $\hbar$ and use
  $\kappa$ as a purely classical coupling constant. This is, of course,
  what was done historically. But it is not clear whether it hides the deeper
  meaning of spacetime structure. In the current approach, ${\cal A}_P$ arises
  as a fundamental unit of area and the coupling constant for gravity
  does involve $\hbar$ if ${\cal A}_P$ is taken as a basic constant. 
  Keeping ${\cal A}_P$ independent of $\hbar$ has non trivial implications 
  when one takes the limits $\hbar \to (0,\infty)$ and this issue needs to 
  be analyzed further.
  
  The fundamental postulate we use is in equation (\ref{postulate}) and it does
  {\it not} refer to any horizon. To see how this comes about, consider any
  spatial plane, say the $y-z$ plane, in flat spacetime. It is always possible to 
  find a Rindler frame in the flat spacetime such that 
  the chosen surface acts as the horizon for some Rindler observer.
  In this sense, any plane in flat spacetime must have an entropy per
  unit area. Microscopically, I would expect this to arise because of the 
  entanglement over length scales of the order of $\sqrt{{\cal A}_P}$.   
  We have defined in (\ref{postulate}) 
  the  entropy per unit area rather than the total entropy in order
  to avoid having to deal with global nature of the surfaces (whether the surface is 
  compact, non compact etc.). 

   An argument is sometimes advanced that Rindler  (or de Sitter) horizon is conceptually different
   from, say, black hole horizon because the former is observer dependent. I believe this argument is 
   incorrect and that all horizons (including even Rindler horizon) should be treated on par
   because of at least three reasons.\cite{tprealms}
   (i) To begin with, if the notion of entropy in black hole
    spacetimes is not accidental,  then
    one would expect {\it any} one-way-membrane which blocks out information to lead to a 
    notion of entropy. (ii) As regards  observer dependence, even in the case of Schwarzschild spacetimes,
    it is possible to have observers moving in time-like trajectories inside the event horizon who will
    access part of the information which is not available to the outside observer. It seems unlikely
     that these suicidal observers will attribute the same amount of entropy to the Schwarzschild
    black hole as an observer playing it safe by staying far away from the event horizon. (iii)
   If the notion of entropy associated with a one way membrane arises from
   {\it local} degrees of freedom and Planck scale physics (as in the case of entanglement 
   entropy) then it should be a local construct. 
  This is also in consonance with the spirit that
  all physical phenomena must be local and the fact that principle of equivalence operates
  in a local region. 
  
  This approach also provides a natural explanation as to why the
  gravitational coupling constant is positive. It is positive because entropy and area
  are positive quantities. 
  
   This approach  emphasizes the role of two dimensional surfaces in fundamental physics
once again, which was  noted earlier in the world sheet action for strings and in the quantization of areas in
loop gravity \cite{abhay92}.  A two dimensional surface is the basic minimum one needs to produce
region of inaccessibility and thus entropy from lack of information. When one connects up
gravity with spacetime entropy it is is inevitable that the coupling constant for gravity
has the dimensions of area in natural units. 
The next step in such an approach will be to find the fundamental units by which
  spacetime areas are made of and provide a theoretical, quantum mechanical
  description for the same. This will lead to the proper quantum description of spacetime
  with Einstein action playing the role of the free energy in the thermodynamic limit of the spacetime.

I thank Apoorva Patel, K.Subramanian and J.V.Narlikar  for comments on the manuscript.

    \end{document}